\documentclass[runningheads]{svmult}
\usepackage{rotate}

\usepackage{makeidx}   
\usepackage{graphicx}  
\usepackage{subeqnar}  
\usepackage{multicol}  
\usepackage{physprbb}  
\makeindex             

\def\ltsima{$\; \buildrel < \over \sim \;$}
\def\lsim{\lower.5ex\hbox{\ltsima}}
\def\gtsima{$\; \buildrel > \over \sim \;$}
\def\gsim{\lower.5ex\hbox{\gtsima}}

\newcommand{\Bbo}{B_{\rm {bo}}}	
\newcommand{\mubo}{\mu_{\rm {bo}}}	
\newcommand{\bp}{B_{\rm{s}}}    

\newcommand{\peq}{P_{\rm{eq}}}   
\newcommand{\pep}{P_{\cal{E}\to\cal{P}}}   
\newcommand{\ppa}{P_{\cal{P}\to\cal{A}}}   

\newcommand{\e}{\cal{E}}
\newcommand{\p}{\cal{P}}	
\newcommand{\ac}{\cal{A}}	
\newcommand{\g}{\cal{G}}
\newcommand{\ra}{r_{\rm{A}}}     
\newcommand{\rcr}{r_{\rm cor}}    
\newcommand{\rg}{r_{\rm G}}      
\newcommand{\rlc}{r_{\rm{lc}}}   
\newcommand{\rns}{R_{\rm{s}}}   
\newcommand{\rst}{r_{\rm{st}}}   

\newcommand{\vrel}{ v_{\rm {rel}}}	

\begin{document}
\title*{SPIN AND MAGNETISM IN OLD NEUTRON STARS}
\toctitle{SPIN AND MAGNETISM IN OLD NEUTRON STARS}
\titlerunning{Spin and magnetism in old neutron stars}
\author{Monica Colpi\inst{1}
\and Andrea Possenti\inst{2}
\and Sergei Popov\inst{3}
\and Fabio Pizzolato\inst{4}}
\authorrunning{Monica Colpi et al.}
\institute{
Universit\`a di Milano Bicocca, Dip. di Fisica, P.zza della Scienza 3,
20133 Milano
\and Osservatorio Astronomico di Bologna, Via Ranzani 1, 40127 Bologna
\and Sternberg Astronomical Institute, Universitetskii Pr. 13, 119899,
Moscow 
\and Istituto di Fisica Cosmica G. Occhialini, Via Bassini 15, 20133 Milano}
\maketitle

\begin{abstract}
The thermal, spin and magnetic evolution of neutron stars in the old low
mass binaries is first explored.  Recycled to very  short periods
via accretion torques, the neutron stars lose their magnetism progressively. 
If accretion proceeds undisturbed for 100 Myrs these stars can 
rotate close to break up with periods far below the minimum observed 
of 1.558 ms. We investigate their histories using population synthesis 
models to show that a tail should exist in the period distribution 
below 1.558 ms. The search of these ultrafastly spinning 
neutron stars as pulsars can help discriminating  among the various equations of state for nuclear
matter, and  can shed light into the physics of binary evolution.

The evolution of isolated neutron stars in the Galaxy is explored
beyond the pulsar phase. Moving through the tenuous interstellar medium, 
these old solitary neutron stars lose their rotational energy.
Whether also their magnetism fades is still a mystery. A population 
synthesis model has revealed that only a tiny fraction of them is able to
accrete from the interstellar medium, shining in the X-rays. There is the
hope that these solitary stars will eventually appear as faint sources 
in the {\sl Chandra} sky survey. This might give insight on the long term 
evolution of the magnetic field in isolated objects.  
\end{abstract}

\section{Introduction}

The amount of rotation and the strength of the magnetic field 
determine many of the neutron star's observational properties.
Over the neutron star lifetime, the spin and the field change and 
the study of their evolution  provides important  clues  
into the physics of the stellar interior.

Of the billion neutron stars in the Galaxy, we shall be mainly concerned 
with the evolution of two distinct populations: The millisecond pulsars 
and the isolated neutron stars both aging, the first in low mass binaries, 
the second as field stars moving in the interstellar medium of our Milky Way.  
We will show that the interaction with their surroundings may profoundly 
alter their spin and magnetic field.  The extent of these changes and the 
modes vary in the two scenarios. It is in exploring this diversity
that we wish to infer the nature of the equation of state and to provide 
a unified view of field decay. In particular, the existence of 
"unconventional" sources, such as 
{\it sub-millisecond pulsars}, stars rotating close to their break up limit, 
and as {\it solitary neutron stars accreting the
interstellar medium,} is a crucial test for our studies.
Their discovery is an observational and theoretical challenge.

In Section 2  we trace the evolution of a neutron
star in the Ejector, Propeller and
Accretion phases, described briefly using simple background
arguments. Section 3 surveys the physical models for the evolution of the
magnetic field both in isolated and accreting systems. In
Section 4 we explore possible individual pathways that may lead to the
formation, in low mass binaries, of neutron stars 
spinning very close to their mass shedding threshold, 
a limit sensitive to the equation of state for nuclear matter.
Pathways of isolated neutron stars follow. 
The first four sections set the general background used to construct,
in Sections 5 and 6, statistical models aimed at determining the
presence and abundance of  sub-millisecond pulsars in the Galaxy,  and of
solitary neutron stars shining in the X-rays.
Specifically, in Section 5 we explore, within the recycling scenario, the star's
spin and magnetic evolution using physical models and 
the role played by disc instabilities in affecting 
the latest phases of binary evolution toward the Ejector
state of sub-millisecond pulsars.
In Section 6 we carry out the first stellar census. We then 
establish how elusive neutron stars can be as accreting sources 
from the interstellar medium due to their large velocities and to
magnetic field decay.
 
\section{The Ejector, Propeller and Accretion Phases}

Over the stellar lifetime,  
magnetic and hydrodynamic torques acting on the neutron star
(NS hereafter) induce secular changes in its spin rate.
Four physical parameters determine the extent of the torques:
the magnetic field strength $B,$ the rotational period $P,$ the 
density of the surrounding (interstellar) medium $n,$ 
and the rate of mass inflow $\dot M$ toward the NS.
According to the magnitude of these quantities, a NS experiences
three different evolutionary paths: Ejector ($\cal E$), 
Propeller ($\cal P$), and Accretion ($\cal A$) \cite{lip}.
In phase $\cal E$ the NS braking results from the loss of 
magneto-dipole radiation as in an ordinary pulsar.
The implied spin-down rate is

\begin{equation}
{d {P}\over dt}={{2\pi^{2}}\over{c^{3} I}}{\mu^2\over P}~~,
\end{equation}

\noindent
where $I$ 
is the NS moment of inertia and $\mu={{\bp \rns^3}/2}$
the NS magnetic moment, function  of the stellar radius 
$\rns$ and polar magnetic field $\bp.$ 
The torque decreases  with increasing $P$, and at the current period
$P$ (much longer than the initial period) the time spent in phase $\e$ is 
\begin{equation}
{\tau_{\e}}={{c^{3}\,I\,P^{2}}\over{4\pi^{2}\,\mu^{2}}}.
\end{equation}
Resulting from the emission 
of electromagnetic waves and charged particles, the rotational 
energy loss of equation (1) can proceed well beyond the active 
radio-pulsar phase and
the magneto-dipolar outflow creates a hollow cavern nesting the NS. 
Phase $\e$ remains active as long as the characteristic 
radius  $\rst$ of this cavern (the {\it stopping radius} \cite{lip} 
\cite{sh71}) is larger 
than both the {\it light cylinder} radius 

\begin{equation}
\rlc~=~{{cP}\over{2\pi}}
\label{lcyl}
\end{equation} 

\noindent
and the {\it gravitational radius}

\begin{equation}
\rg~=~{{2 G M}\over{c_{s}^{2} + v_{\rm rel}^{2}}}
\end{equation}

\noindent
where $c_{s}$ is the sound speed of the interstellar medium (ISM) and $v_{\rm rel}$
the NS velocity
relative to the medium. Phase $\cal E$ ends when matter leaks through 
$\rst,$
and this occurs when the period exceeds 
\begin{equation}
P_{{\cal {E}}\to {\cal{P}}}={{2\pi^{3/4}}\over{c \surd{2GM}}}{\mu^{1/2}v^{1/2}\varrho^{-1/4}}
\label{ep}
\end{equation}
where  $v=(c_s^2+\vrel^2)^{1/2}$ and  $\varrho$ is the mass density of the ISM \cite{tcl93}.
At $P_{{\cal {E}}\to {\cal{P}}}$ the outflowing pulsar momentum flux is unable to balance
the ram pressure of the gravitationally captured matter at $\rg$ \cite{tcl93}.
Matter initiates its infall 
and approaches the {\it magnetospheric radius} (or Alfven radius)  

\begin{equation}
\ra~=~{\left(\mu^{4} \over{2 G M {\dot M}^{2}}\right)}^{1/7}.  
\label{alf}
\end{equation}
\noindent
Here the magnetic pressure of the dipolar stellar field ($\propto r^{-6}$) 
balances the ram pressure of the infalling material ($\propto r^{-5/2}$),
accreting at a rate $\dot M$.
At  $\ra$ the steeply rising magnetic field  would thread  the flow, enforcing it to
corotation.
However, not always can matter pass this edge: 
penetration is prevented whenever the rotational speed $\Omega$ of the 
uniformly
rotating magnetosphere  exceeds the local Keplerian velocity 
$\omega_K$ ($=[GM/\ra^3]^{1/2}$) at $\ra.$
This condition translates into a comparison between 
the {\it corotation radius} 

\begin{equation}
\rcr~=~{\left({G M}\over{\Omega}^{2}\right)}^{1/3}
\end{equation}

\noindent
and $\ra$. When $\rcr$ lies inside $\ra$ the magnetosphere centrifugally 
lifts the plasma above its escape velocity, inhibiting its further infall
toward the NS surface: this is the propeller phase $\cal P$. 
The magnetically driven torques lead to a secular spin-down of the NS 
occurring at a rate

\begin{equation}
{d\over dt}(I \Omega)~=~\xi {\dot M}{\rcr^{2}} \left[\omega_{\rm
K}(\ra)~-~\Omega \right], 
\end{equation}

\noindent
where $\dot {M}$ is the accretion rate at $\ra$, 
and $\xi$ a numerical factor dependent on the accretion pattern, 
ranging from $\xi \simeq 1$ for
disc accretion to $\xi \simeq 10^{-2}$ for spherical accretion.
The effectiveness of the propeller in spinning down the NS  and
in ejecting matter far out from the magnetosphere is still largely
model dependent \cite{lrb-k99}. Recently,
evidence that this mechanism is at work in binaries of
high \cite{swr86} and low mass \cite{c98} has come from X-ray observations.

Phase $\cal P$ terminates only when the corotation radius $\rcr$  
increases above $\ra$; thereon the NS is able to accept matter directly onto 
its surface: this is the accretion phase $\cal A$. The transition 
$\cal P \rightarrow \cal A$ occurs at a critical NS spin rate 
$\ppa$ obtained by equating $\ra$ and $\rcr$:

\begin{equation}
\ppa
=~{{2\pi}\over{\surd{GM}}}
{\left(\mu^4\over{2 G M{\dot M}^{2}}\right)}^{3/14}
\propto ~ {B^{6/7}\over{{\dot M}^{3/7}}}.
\end{equation}

In isolated NS, the accretion flow is almost spherically symmetric and the NS period
$P,$ after the onset of $\ac,$ 
may vary erratically
due to the positive and negative random
torques resulting from the turbulent ISM \cite{lp95} \cite{6kp97}
\footnote{Old accreting isolated NSs are thus expected to show strong fluctuations 
of $\dot P$ over a time scale comparable to the crossing time 
$\tau_{\rm{fluct}}\sim \rm{min}(r_{\rm {ISM}},r_G)/\vrel,$ where
$r_{\rm{ISM}}$ is the spatial scale of the ISM inhomogeneities.}

In a disc-like geometry (in binaries) the spin evolution during accretion 
is guided by 
the  advective, magnetic and viscous torques present in the disc. Depending on
their relative magnitude, these torques can either spin the NS up or down.
The rate of change of the NS angular momentum is obtained integrating 
the different contributions over the whole disc, yielding  
\begin{equation}
{d\over dt}(I \Omega)~\simeq~
{\dot M}\ra^{2}\omega_{\rm K}(\ra)~+~
\int_{\ra}^{\infty}{
B_{z}^{2}{{R^{3} h}\over \eta}{\left[\omega_{\rm K}(R)- \Omega \right]}
}dR~~, 
\label{accspin}
\end{equation}
\noindent
where $\eta$ is the plasma electrical diffusivity,
$h$
the disc
scale thickness and $R$ the radial cylindrical coordinate 
(in eq. \ref{accspin} we neglect the local viscous
contribution as it vanishes at $\ra$). 
The first term in the rhs of equation (\ref{accspin}) describes the advection 
of angular momentum by accretion.
Magnetic torques are instead non-local, resulting from an  integral
that extends over the whole disc: 
different  regions can give either positive (spin-up) or negative
contributions (spin-down). 
When $\Omega \gg \omega_K(\ra),$  
the integrand in (\ref{accspin}) is negative 
yielding a spin-down magnetic torque.
On the other hand, if the star rotates very slowly 
($\Omega\ll \omega_K(\ra)$) the  overall magnetic torque 
is positive,  leading to  a secular spin-up:
when the total torque is positive 
the spin-up process is termed {\it recycling}. 
At a critical (intermediate) value of $\Omega$, termed 
{\it equilibrium angular velocity} $\omega_{\rm eq}$, 
the magnetic torque is negative, and its value exactly 
offsets the (positive) advective torque. In this regime 
the NS accretes matter from the disc
without exchanging angular momentum at all. 
The precise value of $\omega_{\rm eq}$ is rather uncertain, but it ought to lie
between $0.71-0.95$ times the critical frequency $\omega_{\rm cr}=2\pi/\ppa,$
in the mid of the propeller and accretor regimes (see \cite{li99} and references 
therein). Since  $\omega_{\rm cr}$ and $\omega_{\rm eq}$ have near values,
they are often confused in the literature. There is a narrow range 
for $\Omega$ between $\omega_{\rm eq}$ and  
$\omega_{\rm cr}$ 
within which the NS accretes mass but spins down. 

Once the star has reached its equilibrium period 
$\peq=2\pi/{\omega_{\rm eq}}$, further changes 
in the spin period $P$ are possible only if the magnetic field and/or 
$\dot {M}$ vary. Generally, field 
decay causes the star to slide along the so called {\it spin-up line} 
where the period $P$ takes the value $\peq(\mu,\dot {M}),$ at the current values of
$\mu$ and  $\dot {M}.$ 
A large increase  in the mass transfer rate 
can induce further spin-up since the magnetosphere at $\ra$ is squeezed in a region
of
higher Keplerian velocity. 
A decrease in $\dot {M}$ has the opposite effect and 
if there is a decline in  the mass transfer rate (due to
disc instabilities and/or perturbations in the atmosphere of the donor 
star),
the NS can 
transit from $\cal {A}\to\cal {P}\to \cal {E}$.
The critical period for the last transition does not coincide with 
equation (\ref{ep}), but occurs
when the Alfven radius $\ra$ (function of the mass transfer rate $\dot M$)
exceeds  the
light cylinder radius
$\rlc$ (eqs. \ref{lcyl} and \ref{alf}), i.e., when 
\begin{equation}
P_{{\cal{P}}\to {\cal{E}}}={{{2\pi}\over{c}}\left({\mu^{4}\over{2GM\dot M^{2}}}\right)^{1/7}},
\end{equation}
(note that this transition is not symmetric relative to $\cal{E}\to\cal{P}$).
Over the stellar lifetime, NSs  can trace 
loops moving through the various phases 
($\cal {P}\leftrightarrow\cal{A}$ or/and $\cal
{P}\leftrightarrow\cal{E}$). 
It may happen that $\ra$ increases above $\rg$ when 
$\dot {M}$ is very low (as in the ISM):
in this case  the star is in the so called Georotator state ($\g$).

The major drivers of the NS evolution are 
magnetic field decay (though amplification is an interesting possibility)
and variations in the mass transfer
rate: their study is thus 
the subject of $\S 3$ and $\S 4.$

\section{Magnetic field evolution} 

In the  $\mu$ versus $P$ diagram of pulsars (PSRs) we clearly find 
NSs endowed with intense magnetic fields (the canonical pulsars) and 
NSs with a much lower magnetic moment,
the "millisecond" pulsars (MSPs).
The first are isolated objects and largely outnumber the millisecond 
pulsars often living in binaries with degenerate companion stars. 
The origin of this observational dichotomy seems understood but the
physical mechanisms driving the field evolution remain still uncertain.
We review here current ideas for the evolution of the $B$ field. They
are applied later when investigating (i) the existence of sub-millisecond 
pulsars, and (ii) the nature of the six 
isolated NSs discovered by {\sl Rosat}.

\subsection{Historical and observational outline}

Soon after the discovery of PSRs, Ostriker \& Gunn \cite{3og69} proposed 
that their surface magnetic field should not remain constant 
but decay. Under this assumption they explained the absence of 
PSRs with periods much longer than one second (the decay of $\mu$ 
would bring the objects below the death line, turning off the active radio 
emission \cite{3rs75}) and the claimed dimming of the radio luminosity 
$L_R$ proportional to $\mu^2$ \cite{3lmt85}. Their pioneering statistical 
study suggested that the NSs at birth have very high magnetic moments 
$\mu \simeq 10^{29}\div 10^{31}$ G~cm$^3$, whose values decay exponentially 
due to ohmic dissipation in the NS interior, on a typical time-scale \cite{3l1883} 

\begin{equation}
{\tau}_\mu = 4\sigma \rns^2/{\pi}c^2~,
\label{eq_tau-mu}
\end{equation}

\noindent
where $\sigma$ is the conductivity, taken as uniform. Ostriker \& Gunn 
evaluated $\tau_\mu$ using the electrical conductivity for the crystallized 
crust ($\simeq 0.6 \times 10^{23}$ sec$^{-1}$ \cite{3c69}), 
thereby assuming implicitly that the magnetic field resides  
in the outer layers ($\rho < 10^{14}$ g cm$^{-3}$) of the star, and obtained ${\tau}_\mu \sim 4
\times 10^6~{\rm yr}.$
But in the same year, Baym, Pethick \& Pines \cite{3bpp69} argued that 
the magnetic field pervades the entire star and so one had to adopt 
the much higher $\sigma$ of the core, thus predicting a very long $\tau_{\mu}$ 
exceeding the Hubble time, a result confirmed \cite{3cg71} in 1971 when 
different conductivities in the NS interior were taken into account.
When a larger number of pulsars became available for statistics, many authors
compared the observations with  population synthesis Monte Carlo models.
At first \cite{3no90} data seemed to support evidence for field decay,
but subsequent  investigations \cite{3b92} \cite{3w92} \cite{3l94} \cite{3hvbw97} have
indicated that the observed PSR population is
compatible with no decay or with decay  on time-scales ${\tau}_\mu > 10^8$ yr 
longer than the pulsar phase, beyond which the NSs become practically 
invisible.
There is a way to probe indirectly the possible decay of the field
by searching for those old NSs moving with low speed ($<40$ km s$^{-1}$)  
that can gravitationally capture the interstellar medium and shine
as dim accreting sources. As discussed in $\S 6$, their presence 
in the Galaxy depends on the evolution of the field. 

As far as the old  MSPs are concerned, these sources possess 
very low values of $\mu\sim 7\times 10^{25}-10^{27}~{\rm G\,cm^3}$ \cite{3t99}
when they appear as pulsars. The observations 
are again consistent with no field decay over their radio active phase, 
a result inferred from the old age of the dwarf companion stars, 
that implies a $\tau_{\mu}$ at least comparable to their true galactic age
\cite{3bs86}\cite{3k86}\cite{3vvt86}\cite{3vwb90}. This does not preclude 
from the possibility that field decay has occurred during their previous 
evolution: this is discussed and described in $\S 3.4$.

In the next sections we survey the models for the decay of the magnetic field
both for isolated NSs and for the NSs in binaries.

\subsection{Field evolution in isolated objects: spontaneous decay?}
\label{subsec:sdd}

Models for magnetic field decay in isolated NSs flourished over the 
last decade. A first improvement was to abandon the hypothesis of a uniform 
conductivity in the stellar crust and core; $\sigma$ increases when moving 
from the outer liquid layers to the deeper crystallized crust \cite{3yu80}. 
Sang \& Chanmugam \cite{3sc87} noticed that the decay of a field
residing only in the crust is not strictly exponential because of its inward 
diffusion toward regions of higher $\sigma.$ Moreover, the decay rate depends 
on the (highly uncertain) depth penetrated by the initial field:
the decay is more rapid if, at birth, the field resides in the outer lower 
density regions of the crust both because $\sigma$ is lower and because 
$\tau_{\mu}$ depends on the length scale of field gradients (less than a 
tenth of the stellar radius $R$ if $\mu$ is confined in the crust). 
  
The inclusion, in the calculation, of the cooling history of the NS was a 
second major improvement \cite{3sct90} \cite{3um92} \cite{3uk98}. 
It confirmed the non exponential behavior of the decay, showing in addition 
that a slower cooling would accelerate the decay (as a warmer star has a 
lower conductivity). A first short phase ($\lsim 10^6$ yr) of comparatively
rapid decay was found, during which $\mu$ drops from values 
$\sim 6\div30 \times 10^{30}$ G\,\,cm$^3$, typical of young pulsars in 
supernova remnants, to more canonical values $\sim 1\div3 \times 10^{30}$ 
G cm$^3$ typical of normal PSRs. (During this phase the electron-phonon 
interaction dominates $\sigma$). A second phase of no decay was found 
lasting $10\div300$ Myr, followed  by a power-law decay 
(dominated by the interaction of the electron upon the impurities)
and eventually an exponential decay.

If the field penetrates the whole star, the core ohmic decay 
time-scale would be too long to allow  decay \cite{3sp95}. 
Thus, an alternative scenario was proposed  on the hypothesis 
that NSs have  superfluid and superconducting cores. The angular 
momentum of the core  is believed to be carried by \cite{3r72} 
$N_{\rm{vortex}}~\simeq~2~\times~10^{16}/P_{\rm sec}$
neutron vortex lines, parallel to the spin axis; 
the magnetic flux should be confined in $N_{\rm{fluxoid}}$ quantized 
proton flux tubes, where $N_{\rm{fluxoid}}~\simeq~10^{32}~B_{C,13}$,
with $B_{C,13}$ the average field strength in the core, expressed in 
units of $10^{13}$ G. In 1990, Srinivasan {\it et al.} \cite{3s90} 
suggested that the inter-pinning \cite{3mt85} between these quantized 
entities causes the fluxoids to be carried and deposited in the crust 
as the NS spins down. The magnetic flux penetrated into the crust will 
then decay due to ohmic dissipation and the surface magnetic field of the 
NS keeps decreasing until it relaxes at the value of the residual field 
in the core. In this framework the {\it spin history} of the star drives 
and controls the changes of the magnetic field. Assuming the same radial 
velocity for both the fluxoids and the vortices this model implies 
a time-evolution

\begin{equation}
\mu(t)=~\mu_0~\left(~1~+~{t}/{\tau_{\rm {sdd}}} \right)^{-1/4}
\label{eq_sif_simple}
\end{equation}

\noindent
occurring on a spin down time 
\begin{equation}
\tau_{\rm {sdd}}
\sim 8\times10^6~~{P_0^2~I_{45}}{\mu_{0,30}^{-2}}~~{\rm yr},
\end{equation}

\noindent
where index $0$ denotes values at the onset of phase $\cal E$, and 
$I_{45}$ the NS moment of inertia in units of $10^{45}$ g cm$^2$.
Equation (\ref{eq_sif_simple}) models the simplest version of the
{\it spin down driven} (sdd) decay of $\mu$ in isolated NSs. It can
be refined, accounting for the non instantaneous relaxation of the
surface magnetic field to the value in the core \cite{3mb94}.

Disappointedly, until now it has not been possible to perform a reliable statistical 
test of all the models. They predict clearly distinct evolutionary 
pathways only during the early stage of the pulsar life, namely during 
the first $10^6$ yr after birth. Therefore a large sample of young NSs
is requested for a comparison. Up to now, the pulsar catalog lists only 
an handful of such objects, as the observed PSR population is dominated 
by elder sources, with characteristic time of few $10^7$ yr. In such 
situation, there is no statistical clue for rejecting the hypothesis of 
a non decaying field. Some of the surveys in progress, as the one running 
at Parkes \cite{3c00}, dedicated to the search of young PSRs, could 
ultimately solve this problem.

\subsection{Magnetic field evolution in binaries: accretion driven scenario}

None of the models for spontaneous decay, proposed so far
for the isolated objects, can explain magnetic moments of 
$\sim 10^{26}$ G cm$^3$, which are typical of the millisecond PSRs.
In order to attain these values of $\mu$ over  a Hubble time, 
the crustal models would require currents located only in a 
narrow layer below a density $\rho < 6\times 10^{10}$ gcm$^{-3}$, a 
very unlikely possibility. On the other hand, the "sdd" models would  
predict only a modest decrease of $\mu,$ at most of two orders of magnitude 
with respect to the initial values. However, one common feature groups 
the low-$\mu$ objects, i.e., that {\it all the PSRs with low magnetic 
moment spent a part of their life in a multiple stellar system}.
Guided by this fact, many authors have tried to relate the low values of $\mu$
with the interaction between the NS  and its companion star(s). 
Two different scenarios have been proposed so far:
(i)  The {\it accretion driven scenario} developed under the hypothesis 
of a {\it crustal magnetic field};
(ii) The {\it spin driven scenario} lying on the hypothesis of 
a {\it core magnetic field}.

In this section we consider scenario (i) taking the hypothesis that
the currents generating the field are located deep in the crust 
\footnote{
The onset of a thermo-magnetic instability, which transforms heat into 
magnetic energy at the moment of NS formation, is an effective mechanism 
to produce strong fields in the crust of a NS \cite{3uly86} \cite{3wg96}. 
Although this instability is not yet completely explored for poloidal fields, 
it is a plausible mechanism for the origin of a crust field which 
does not depend on special assumptions.}.
A first observational suggestion about the effects of the accretion on a 
(crustal) magnetic field was given in 1986 by Taam \& van den Heuvel 
\cite{3tv86}: they noticed an approximate dependence of the surface magnetic 
field  on the amount of mass accreted onto the neutron star.
Shibazaki {\it et al.} \cite{3s89} later  presented an empirical
formula for the decay in presence of accretion:

\begin{equation}
\mu~=~\frac{\mu_0}{1~+~{\Delta M_{\rm {acc}}}/{m_*}}
\label{eq_shibazaki}
\end{equation}

\noindent 
where $\mu_0$ represents the magnetic moment at the onset of accretion, 
$\Delta M_{\rm {acc}}$ is the amount of accreted mass and $m_*$ a parameter
to be fitted with the observations. These authors claimed that $m_*\sim 10^{-4}
{\rm {M_\odot}}$ could reproduce the Taam \& van den Heuvel's correlation.
Zhang, Wu \& Yang \cite{3zwy94} \cite{3z98} gave physical foundation to 
(\ref{eq_shibazaki}) assuming that the compressed accreted matter
has ferromagnetic permeability. However, using a larger database 
Wijers \cite{3w97} showed that these  models are not fully consistent 
with the available data both on X-ray binaries and recycled pulsars. 
Romani \cite{3r90} first introduced the accretion rate $\dot{M}$ 
as a second parameter in driving the magnetic moment decay
in addition to $\Delta M_{\rm{acc}}$. He pointed out
that the accretion produces two major effects: (i) heating of the crust
(depending on $\dot{M}$), which determines a reduction of the
conductivity (and in turn the hastening of ohmic decay) 
and (ii) advection of the field screened by the diamagnetic accreted 
material. 
Moreover, the advection of the
field lines stops when $\mu \lsim 10^{27}$ G cm$^3$, resulting in an
asymptotic value for the magnetic moment (in agreement with the observations).

The existence of a bottom field (with a predicted scaling 
$\mu \propto \dot{M}^{1/2}$) is also possible if 
the currents sustaining the surface magnetic field of the NS are 
neutralized by the currents developed in the diamagnetic blobs of 
accreted matter \cite{3al80}. In this case the decay of $\mu$ typically ends
when the accretion disc skims the NS  surface \cite{3bkw96}.

The first fully consistent theoretical calculations of the 
accretion-induced-decay of the crustal magnetic field of a NS  was  
performed by Urpin, Geppert \& Konenkov 
\cite{3gu94} \cite{3ug95} \cite{3ug96} \cite{3ugk98a}.
The basic physical mechanism is the diffusion (ohmic decay of currents) 
and advection of the magnetic field sustained by currents circulating 
in the non superconducting crust. The magnetic field 
evolution is calculated according to the induction equation:
\begin{equation}
\frac{\partial \vec{B}}{\partial t} = - \frac{c^{2}}{4 \pi}
\nabla \land \left( \frac{1}{\sigma} \nabla \land \vec{B}
\right) + \nabla \land ( \vec{v} \land \vec{B} ) 
\label{eq_induction2}
\end{equation}
\noindent
where $\vec{B}$ is the magnetic induction and $\vec{v}$ is the
velocity of the moving fluid ($=0$ in a non accreting NS,
$|\vec{v}|=\dot{M}/4\pi r^2\rho$ for a radial approximation
of the accreted fluid flow). 
In their calculation, the superconducting NS core 
is assumed to expel the magnetic field (in the following we will refer to this 
boundary condition for the field at the crust-core interface as 
BCI); the induction equation is solved for a dipolar field
and $\sigma$ is given as in \cite{3p00}. 
\begin{figure}[htbp]
\begin{center}
\includegraphics[width=0.65\textwidth]{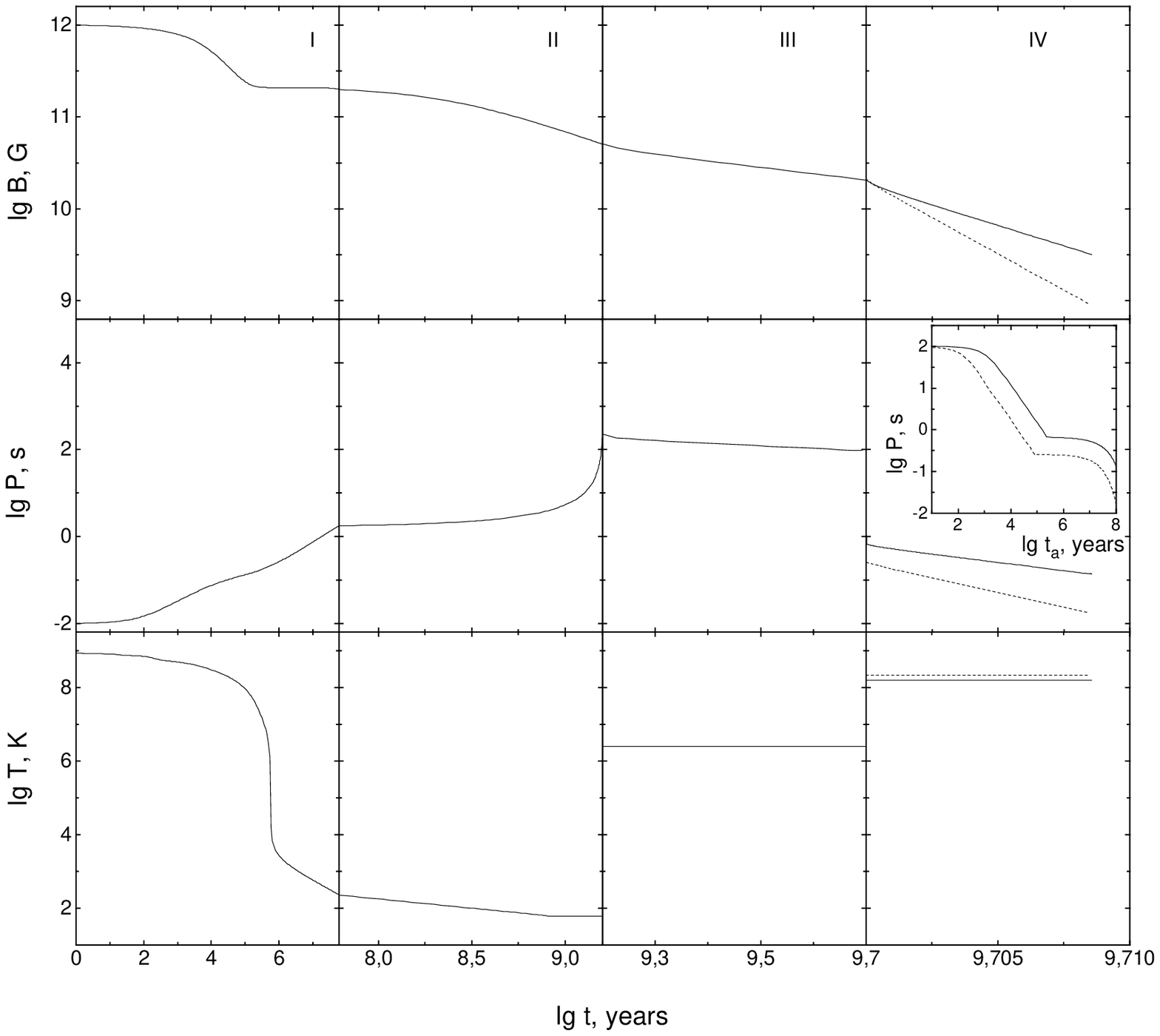}
\end{center}
\caption[]{Magnetic, rotational and thermal evolution vs time of a 
$1.4~{\rm M_\odot}$ NS with  $\mu=10^{30}$ G cm$^{3}$ 
at birth, as calculated  in \cite{3ugk98}, under BCI. The adopted equation 
of state is PS (a stiff
one). The initial depth of the currents is $10^{13}$ g cm$^{-3}$ and the 
impurity parameter $Q=0.01$. 
Phase I and II refers to $\e$ and $\p$ respectively. 
The value of $\dot {M}$ from the stellar wind 
(referred to as phase III of $\ac$) is $2/3 \times 10^{-8} {\rm \dot{M}_{E}}$, and the
lifetime 
of the companion star on the main sequence is $5\times 10^9$ {\rm yr}. During 
the disc accretion phase (referred to as phase IV), $\dot {M}$  is $2/3 \times
10^{-3} {\rm \dot{M}_{E}}$
({\it solid line}) or $2/3 \times 10^{-2} {\rm \dot{M}_{E}}$
({\it dotted line}). Phase IV lasts $10^8$ yr. The inserted panel
enlarges the spin evolution during the very short phase at the boundary
between phase III and IV, that is when disc accretion sets in.
}
\label{fig_gepp_dec}
\end{figure}

Accretion affects the ohmic decay
in two ways: (i) {\it heating the crust,} so reducing $\sigma$;
(ii) {\it transporting matter and currents} toward the core-crust boundary.
The relative importance of these two effects depends on other physical
parameters, such as the initial depth penetrated by the currents, the
impurity content ($Q$), the accretion rate, the duration of the accretion phase
and the equation of state of the nuclear matter.
In coupling the magnetic history of the star to its spin evolution
(through $\e$ to $\ac$), 
Urpin, Geppert \& Konenkov \cite{3ugk98} found that 
{\it neutron stars born with standard magnetic moments
and spin periods can evolve to low-field rapidly rotating objects},
as illustrated in  Figure
\ref{fig_gepp_dec}.

The behavior of currents and the response of nuclear matter to
an interior field, at the interface between the crust and the
core of a NS (typical density $\simeq 2\times 10^{14}$ g cm$^{-3}$),  
is still an open issue for the theorists. In view of these
uncertainties, Konar \& Bhattacharya \cite{3kb97} chose a different 
boundary condition at the crust-core interface for solving the 
equation (\ref{eq_induction2}). They noticed that the deposition
of accreted matter on top of the crust can imply the assimilation of original
current-carrying crustal layers into the superconducting core of the 
NS. In particular they assumed \cite{3kb97} the  
{\it newly formed superconducting material to retain the  magnetic 
flux} in the form of Abrikosov fluxoids \cite{3bpp69} rather than to expel it
through the Meissner effect (in the following
this boundary condition will be labeled as BCII). Within this model, 
accretion produces a third effect on the evolution of $\mu$:
(iii) the {\it assimilation of material into the core}, 
where the conductivity is huge, freezes the decay.
\begin{figure}[htbp]
\begin{center}
\includegraphics[width=.47\textwidth]{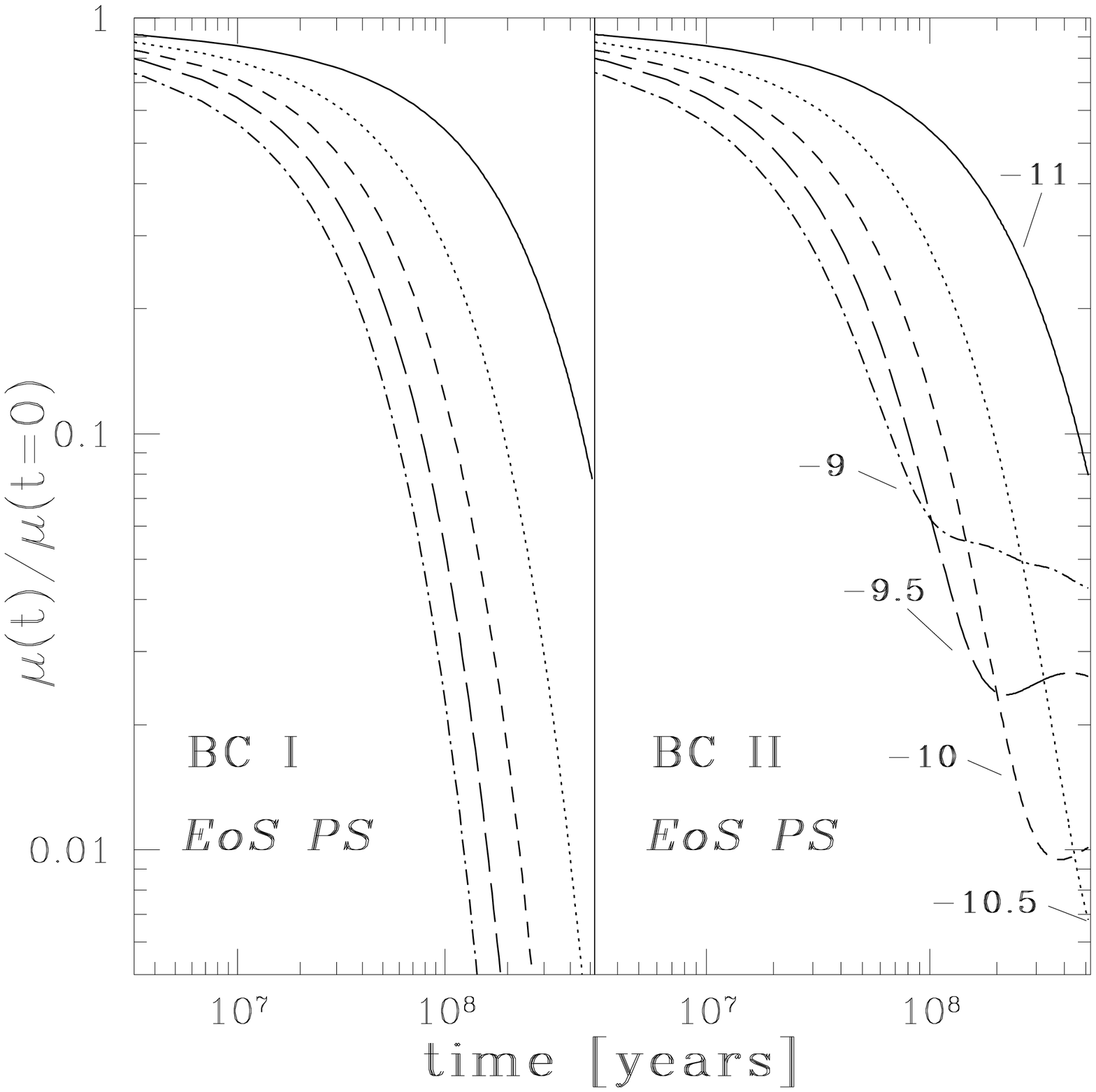}
\end{center}
\caption[]{Evolution, under 
BCI (left) and BCII (right panel), of the surface magnetic moment 
$\mu(t)$ (relative to its initial value) of a $1.4~{\rm
M_\odot}$
NS undergoing accretion at five different levels. Top (bottom) curve refers to  
 $\dot{M}=10^{-11}$ M$_{\odot}$  y$^{-1}$ ($\dot {M}=10^{-9}$ M$_{\odot}$ y$^{-1}$),
from 
\cite{5pcgbd99}. Equation of state "PS" is adopted.}
\label{fig_kb_dec}
\end{figure}

As illustrated in Figure \ref{fig_kb_dec}, also {\it this hypothesis
leads to a reduction of the surface magnetic field of more than 4 orders
of magnitude} in $10^8$ yr, explaining the low $\mu$ and its possible 
freezing in the millisecond pulsar population. Likewise for BCI,
it emerges that the decay of $\mu$ depends not only on the total accreted 
mass, but on the accretion rate itself. However, in this case 
the higher the accretion rate is, the stronger is the pull of crustal 
material into the core resulting in an earlier  freezing of the decay. 
In a subsequent paper \cite{3kb98}, it was claimed that such a positive 
correlation between $\dot{M}$ and the final value of $\mu$ 
is supported by some observational evidence \cite{3pl97} \cite{3wz97}.

Other refinements of the accretion-induced decay scenario include 
the effects of a non spherical symmetric accretion \cite{3s98},
the evolution of multipoles at the NS surface \cite{3s98},
the post-accretion increase of $\mu$ due to the re-diffusion of the buried 
field towards the surface \cite{3yc95}, and the relativistic corrections
to the decaying history \cite{3sen98}. All together, they produce only
slight changes to the aforementioned results.

\subsection{Magnetic field evolution in binaries: spin driven scenario}

Both the simple idea of the flux conservation during the gravitational collapse
and the action of a dynamo in the convective proto-neutron star \cite{3td93}
lead to the existence of a magnetic field penetrating the whole NS.
Due to the huge conductivity of the core matter, 
no ohmic decay occurs during a Hubble time and so other effects have been 
invoked to account for the low field NSs. 

Besides magneto-dipole braking, a NS experiences a longer and more significant 
phase of spin-down, phase $\cal P$ (described in $\S 2$) 
which enhances the effect of the spin-down driven scenario sketched in 
\S \ref{subsec:sdd}. This is particularly important when the NS lives 
in a binary system. Miri \& Bhattacharya \cite{3mb94} and
Miri \cite{3m96} explored the case of low mass systems
showing that at the end of $\cal P$ and $\cal A$ phases 
the magnetic moment $\mu$ has decayed, relative to its initial 
value $\mu_0$, by a factor scaling with $P_0/P_{\rm {max}}$ where
$P_0$ is the initial rotational period and $P_{\rm {max}}$
the longest period attained during the phase where $\cal P$ is active 
($\cal P$ results  from the interaction of the companion wind with the NS).
Assuming that the star spins-down to $P\lsim 10^3$ sec, a residual  
magnetic moment is 

\begin{equation}
\mu_{\rm {final}}~\simeq~\frac{0.1\div 1~{\rm sec}}{1000~{\rm sec}}~\mu_0
~\simeq~10^{-4}\div10^{-3}~\mu_0
~\simeq~10^{26}\div 10^{27}~{\rm G\,\,cm^{3}}
\label{eq_sif_mufinal}
\end{equation}

\noindent
compatible with the values observed for MSP population.

Recently, Konar \& Bhattacharya \cite{3kb99} re-examined this scenario
in the case of accreting NSs, incorporating the microphysics of the crust 
and the material movements due to the accretion. They concluded that the
model can reproduce the values of $\mu$ observed in the low-mass systems only
for large values of the unknown impurity parameter $Q\gsim 0.05$ (in contrast
with accretion driven decay models demanding for values of  $Q\lsim
0.01$). If 
the wind accretion phase is short or absent, $Q$ should even exceed unity. 
A more serious objection to this model was risen by Konenkov \& Geppert 
\cite{3kg99} who pointed out that the motion of the proton
flux tubes in the core leads to a distortion of the field structure near
the crust-core interface and this in turn creates a back-reaction of the crust
on the fluxoids expulsion: it results that (i) the sdd can be adequate 
for describing the $\mu$-evolution only for weak initial magnetic moments 
($\mu_0 \lsim 10^{29}$ G cm$^3$) and (ii) the predicted correlation 
$\mu_{\rm {final}} \propto P_{\rm {max}}^{-1}$ is no more justified.

Finally, we have to mention another class of models in which the
magnetic evolution is strictly related with the spin history of the star:
in 1991 Ruderman \cite{3r91b} \cite{3r91c} proposed that the coupling among
$P$ and $\mu$ could take place via crustal plate tectonics. The rotational
torques acting on the neutron star cause the crustal plates to migrate,
thus dragging the magnetic poles anchored in them. As a consequence the
effective magnetic dipole moment can strongly vary, either in intensity
(decreasing or growing) and in direction, suggesting a tendency to produce
an overabundance of both orthogonal (spin axis $\perp$ magnetic axis)
and nearly aligned rotators (spin axis $\parallel$ magnetic axis).
Current observation of disc population MSPs seem compatible 
with this feature \cite{3crz98}. Moreover the crust-cracking events
could account for the glitches seen in young PSRs \cite{3rzc98}.
A more extended description of the variety of the spin driven decay 
models can be found for example in \cite{3r95}.

\section{Evolutionary pathways in various environment}

We here describe shortly possible NS pathways 
driven by the changes in the field and in the accretion rate, over 
phases $\cal E$, $\cal P$ or 
$\cal A,$ in low mass binaries
and in the ISM.

\subsection{Binaries}

The NSs in binaries experience a  complex evolution
which is tightly coupled to the orbital and internal evolution of
the companion star \cite{4bv91}.
The powerful pulsar wind initially sweeps the stellar wind 
away and the NS spends its lifetime in the $\cal E$ phase.
With the weakening of the electro-magnetic pressure with 
increasing period, an extended $\cal P$ phase establishes (lasting 
$10^{8}-10^{9}$ yr) during which the NS is spun down further to periods 
of $\sim 10^{2}-10^{4}$ seconds in its interaction with the stellar
wind of the companion star. When the period $\ppa$ is 
attained, accretion sets in down to the NS surface and we could 
possibly trace this phase identifying the X-ray emission from the 
wind fed NS. Here the NS can accrete with no exchange of angular 
momentum with the incoming matter unless the magnetic field decreases  
due to either accretion induced or spin induced decay. The NS can slide 
along the corresponding equilibrium spin-up line.
As soon as the donor star evolve into a giant state,
it can fill its  
Roche lobe and matter overflows from the inner
Lagrangian point forming an accretion disc. When a disc establishes
(on the viscous time scale) large accretion rates are available
and from this moment on recycling starts. In the Roche lobe  
overflow phase (RLO) the NS can be re-accelerated to millisecond
periods while the field decays down to values of $10^8-10^9$ G. 
The mass transfer and orbital evolution are complex
to model and it is now believed that a significant fraction
of the mass lost by the donor does not accrete onto the NS, but
is ejected from the system \cite{4ts99}.

\begin{figure}[htbp]
\begin{center}
\includegraphics[width=.47\textwidth]{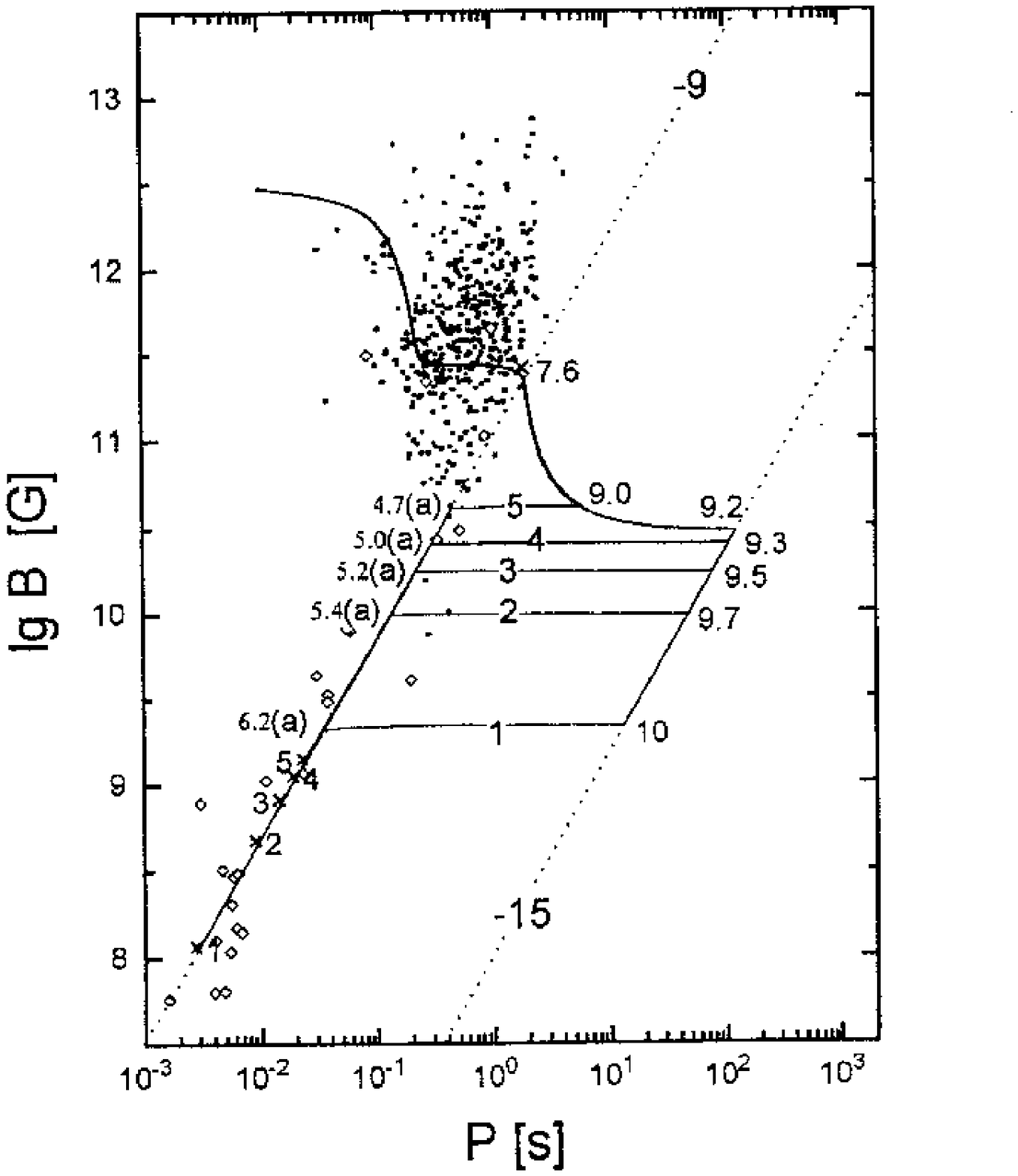}
\end{center}
\caption{The evolutionary tracks of NSs in the plane
$B-P$ for different durations of the evolution (labeled by the numbers 
on the right of the curves, in log$_{10}$ yr) during  phases 
I($\e$) + II($\p$)+ III ($\ac$) + IV($\ac$). The initial surface magnetic field is $3
\times 10^{12}$ G, $Q=0.03$, and the other parameters as in 
Figure \ref{fig_gepp_dec} \cite{3ugk98a}.}
\label{fig_gepp_p-b}
\end{figure}

Once reached the spin-up line at  the current accretion rate, 
the NS loads matter  acquiring angular momentum at the rate 
of field decay.
Recycling ceases when the donor star has evolved beyond the red giant phase
and a dwarf remnant is left over stellar evolution, or when 
the binary as become detached, so that the donor underfills 
its Roche lobe  \cite{4bv91}. If accretion ends abruptly, 
the NS may avoid phase $\cal P$, transiting directly to $\cal E$ and 
possibly re-appearing as an active radio pulsar. In low mass binaries where 
the donor star is a low main sequence star, evolution proceeds in a 
fairly ordered way: from $\cal E$ to $\cal P$ (from the stellar wind), 
to $\cal A$ (from wind fed accretion), to $\cal A$ again 
(from a Keplerian disc; RLO), as shown in Figure \ref{fig_gepp_p-b}.
Eventually the NS transits to $\cal E$ or 
$\cal P/\cal E$ when accretion halts. Observational and theoretical 
considerations hint for a large decay of the magnetic moment $\mu$ 
in phase ${\cal A},$ particularly during 
RLO, as outlined also in Figures 1 and 3.
The type of evolution will be applied in $\S 5$ to address the problem of the
existence of sub-millisecond pulsars.

\subsection{Isolated neutron stars}

At birth, isolated NSs experience phase $\e$. The subsequent evolution depends
on the NS velocity relative to the
ISM, the value of the ISM density  and  the magnetic field intensity. The large
mean kick
velocities acquired at birth 
($\langle V \rangle\sim 300$ km s$^{-1},$ \cite{3l94}
\cite{3hvbw97}\cite{4hp97}\cite{4cc97}) can 
reduce the extent of magnetic torques in phase $\p$ and
can 
impede accretion fully. 
It is thus possible  that
phase $\cal {E}$ never ends. This occurs when the NS speed $V$ is as
high as $V>100\, \mu_{30}\, n^{1/2}$ km s$^{-1}:$ This speed is derived estimating the
duration of phase $\cal {E}$ at $\pep$ (eqs. 2 and 5) 
\begin{equation}
\tau_{{\cal{E}}\to{\cal{P}}}=\tau_{\e}(\pep)\sim 10^9n^{-1/2}V_{10}\mu_{30}^{-1}\rm
{yr}
\end{equation} 
imposing  $\tau_{\e\to\p} \sim 10^{10}$ yr and $v\sim V$ (in eq. 18, 
$V_{10},\,\,n,$ and $\mu_{30}$ are the velocity, density,  and 
magnetic moment in units of 10 km s$^{-1},$ 1 cm$^{-3},$  and 
10$^{30}$ G cm${^3}$, respectively; hereon we will denote the normalizations as
subscripts,
for simplicity). 
Clearly,  only the
slowest NSs can loop into the accretion phase under typical ISM conditions (
where $n\lsim 1$ cm$^{-3}$). 
Molecular clouds seem a privileged site for 
phase $\ac$ \cite{4cct93}, but the probability of crossing these high density
regions is relatively low and shortlived.

\begin{figure}[htbp]
\begin{center}
\includegraphics[width=.5\textwidth]{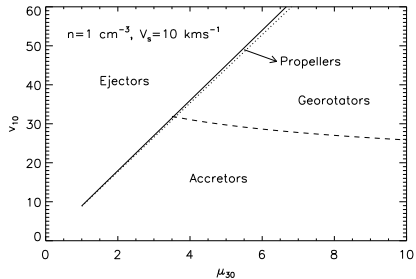}
\end{center}
\caption{Phases $\e,\p,\ac$ and $\g,$ in the $V_{10}-\mu_{30}$ plane for
isolated NSs moving in a ISM
of density $n=1$  cm$^{-3}$ and a sound speed of 10 km s$^{-1}$, from \cite{p2000}.}
\label{Vvsmu_iso}
\end{figure}

Typically, in the ISM accretion is nearly spherical and a Bondi-Holye type flow
establish with
\begin{equation}
{\dot{M}}=2\pi G^2 M^2\rho\left (c_s^2+\vrel^2\right)^{-3/2}
\end{equation}
where here
$\vrel$ coincides with $V$.
The critical periods thus are 
\begin{equation}
\pep\sim 10\,n^{-1/4}V_{10}^{1/2}\mu_{30}^{1/2}\,{\rm {s}}\,\,\,\,\,{\rm
{and}}\,\,\,\,\,
\ppa\sim 500\,\, n^{-3/7}V_{10}^{9/7}\mu^{6/7}_{30}\,\rm{s},
\end{equation}
if the magnetic field is constant in time, $c_s\sim 10$ km s$^{-1}$, and $M=1.4$
M$_{\odot}$.
As indicated in equations (18-20), the NS evolution
is guided by  the two key parameters, $V$ and $\mu$ (at fixed ISM's density). This
is quantitatively illustrated in Figure \ref{Vvsmu_iso} where 
the NS's phases are identified in the $V$ vs $B$ plane.
A  strong constant magnetic field ($\mu>10^{30}$ G cm$^3$)
implies large magneto-dipole losses that decelerate rapidly the
NS to favor its entrance in phase $\cal {A}$ (even when $V\sim 300$ km s$^{-1}$),
despite the
long period of the $\p\to\ac$ transition  
(note eq. 20).

What  are the consequences of field decay on the evolution? 
Whether field decay enhances
or reduces the probability of a transition beyond $\e$ is a subtle question
that has recently been partly addressed 
\cite{4ctzt98}\cite{4lxf98}\cite{4ttzc00}\cite{4pp00}.
Two competing effects  come into play when $B$ decays: 
(i)
the spin-down rate slows down
because of the weakening of $B$, causing the NS to persist longer
in  state $\e;$
(ii) the periods $\pep$ and $\ppa$ instead decrease with $B,$ and this acts in the
opposite
sense.

\begin{figure}[htbp]
\begin{center}
\includegraphics[width=.46\textwidth,angle=-90]{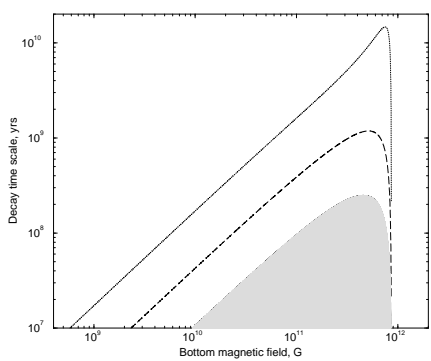}
\end{center}
\caption[]{Loci, in the $\tau_{\mu}-\Bbo$ plane, 
where phase $\e$ lasts longer than 10$^{10}$ yr. 
The curves refer (from top to bottom) 
to initial fields of $5\times 10^{11}$ (dotted line), $10^{12}$(dashed line), $2\times
10^{12}$ G (hatched region), for   $\vrel+c_s=40$ km s$^{-1}$, and
ISM density $n=1$
cm$^{-3}.$ 
}
\label{bdecay_iso}
\end{figure}

To explore  this delicate interplay,  
Popov \& Prokhorov~\cite{4pp00} studied a toy
model where a 
field is exponentially 
decaying on a  scale $\tau_{\mu}$ from its initial value $\mu_0$ down to a bottom value
$\mubo.$  Under this hypothesis, the
Ejector time scale (with a decaying $\mu$) can be estimated analytically to give
 
\begin{equation}
\tau_{{\e},\mu}= \cases {-\tau_{\mu}\ln\left [
\displaystyle{{\tau_{\e}\over \tau_{\mu}}}\left (1+
\displaystyle{{\tau_{\mu}^2\over \tau_{\e}^2}}
\right )^{1/2}-1\right ] & 
$\tau_{{\e},\mu}<\tau_{\rm {bo}}$ 
\cr & \cr
\tau_{\rm {bo}}+ 
\displaystyle{{\mu_0\over \mubo}} \tau_{\e} 
-{1\over 2}\tau_{\mu}{\mu_0\over \mubo} 
\left (1-e^{2\tau_{\rm {bo}}/\tau_{\mu}}\right )  &
$\tau_{{\e},\mu}>\tau_{\rm {bo}}$\,\,\,, \cr}
\end{equation} 
where $\tau_{\rm {bo}}=\tau_d \ln(\mu_0/\mubo)$ is the time when
the bottom field is attained and $\tau_{\e}$ is from equation (2) for a constant
field.
Figure \ref{bdecay_iso} shows the loci where the time spent in phase $\e$ equals
the age of the Galaxy, i.e., $10$ Gyr, as a function of the
bottom field $\Bbo$ and of the decay time-scale $\tau_{\mu}$.
The ``forbidden'' region lies just below each
curve. As it appears from the Figure, the interval over which 
the NS never leaves stage $\e$ is
non negligible: It widens when $\mu_0$ is decreased, due to the weakness of the
magneto-dipole torque. 
Large as well as very low bottom fields do not constrain $\tau_{\mu}$ 
and permit  entrance to phase $\ac$.
When $\mubo$ is very low, there is a turn-off
of all magnetospheric effects on the inflowing matter and matter accretes promptly.

Similarly, 
Colpi {\it et al.} \cite{4ctzt98} considered a model in which the
spin evolution causes the core field to migrate to the crust where
dissipation processes drive ohmic field decay (Sect. 3).
In this circumstance, the entrance to phase $\ac$ is less likely
if  $\tau_{\mu} \sim 10^{8}$ yr, but possible otherwise. 
Thus, field decay can hinder the stars in phase $\e$
if its decay is somewhat "fine-tuned", while    
a fast decay would drive them into $\ac$ promptly. 
The strongest theoretical argument against phase $\ac$
remains however the high kick velocity that the NSs
acquire at birth (see the review of Lai in this book).
There remain nevertheless open the possibility that
weak field accreting NSs exist in the Sun's vicinity.

We now proceed on exploring the existence of "unusal" NSs,
the sub-MSPs and the X-ray isolated NSs for which
these pathways are statistically relevant.   

\section{A population synthesis approach to the formation of {\it sub}-MSPs}

\begin{table}    
\caption{Disc accretion onto a $1.4$ M$_\odot$ neutron star}
\begin{center}
\renewcommand{\arraystretch}{1.4}
\setlength\tabcolsep{5pt}
\begin{tabular}{c||c|c||c|c|c||c|c|c}
\hline\noalign{\smallskip}
EoS                 & 
Fate                & T/W$_{fin}$           &
$\Delta$M$_B$       & M$_{G,fin}$           & $P_{fin}$        & 
M$_{G,max}^{static}$  & M$_{G,max}^{rotating}$  & P$_{min}^{abs}$  \\
\hline\noalign{\smallskip}
  F   & Collapse & 0.034 & 0.172 & 1.52 & 0.72 & 1.46 & 1.67  & 0.47 \\
  A   & MassShed & 0.120 & 0.428 & 1.77 & 0.60 & 1.66 & 1.95  & 0.47 \\
  E   & MassShed & 0.115 & 0.414 & 1.76 & 0.66 & 1.75 & 2.05  & 0.48 \\
  AU  & MassShed & 0.126 & 0.446 & 1.79 & 0.70 & 2.13 & 2.55  & 0.47 \\
  D   & MassShed & 0.111 & 0.405 & 1.76 & 0.73 & 1.65 & 1.95  & 0.57 \\
  FPS & MassShed & 0.114 & 0.416 & 1.76 & 0.75 & 1.80 & 2.12  & 0.53 \\
  UT  & MassShed & 0.119 & 0.429 & 1.78 & 0.75 & 1.84 & 2.19  & 0.54 \\
  UU  & MassShed & 0.121 & 0.436 & 1.78 & 0.78 & 2.20 & 2.61  & 0.50 \\
  C   & MassShed & 0.103 & 0.389 & 1.74 & 0.89 & 1.86 & 2.17  & 0.59 \\
  N*  & MassShed & 0.130 & 0.484 & 1.84 & 1.08 & 2.64 & 3.22  & 0.68 \\ 
  L   & MassShed & 0.116 & 0.443 & 1.80 & 1.25 & 2.70 & 3.27  & 0.76 \\
  M   & MassShed & 0.091 & 0.367 & 1.74 & 1.49 & 1.80 & 2.10  & 0.81 \\
\hline\noalign{\smallskip}
\end{tabular}
\end{center}
\footnotesize{
\noindent
The second and the third columns contain the fate of the
NS at the end point of recycling and the resulting ratio 
of kinetic energy over gravitational energy. The 
fourth and fifth columns 
report the total accreted baryonic mass and the final gravitational mass 
(in units of solar masses). The sixth column lists the final attained 
rotational periods. 
The seventh column collects the values of the maximum mass for a
non-rotating spherical configuration, the eighth for a maximally
rotating star and its corresponding minimum period for stability
(ninth column). The table is from Cook, Shapiro \& Teukolsky
\cite{5cst94}.}
\label{tab:eos}
\end{table}
This section is devoted to the statistical study of 
the old NSs in binaries in the aim at exploring the recycling process
and the possibility that ultra fastly spinning NSs can form in these systems.
The evolutionary scheme is based on  $\S 2, 3.3,$ and $\S 4.1.$
A similar approach is presented in $\S 6$ for the isolated NSs in the Galaxy.

On the observational side, the NS having the shortest
rotational period $P_{\rm{min}}=1.558$ ms ever detected
is PSR~1937+21.
Despite its apparent 
smallness, $P_{\rm{min}}$ is not a critical period for NS rotation: as shown 
by Cook, Shapiro \& Teukolsky \cite{5cst94}, the period $P_{\rm{min}}$ is 
longer than the limiting period, $P_{\rm{sh}}$, below which the star becomes 
unstable to mass shedding at its equator, irrespective to the adopted
equation of state (EoS) for the nuclear matter.
Table~\ref{tab:eos} just shows the values of $P_{\rm{sh}}$ for a set of EoS and
the corresponding values of accreted baryonic mass.
It shows how sensitive is $P_{\rm{sh}}$ to the EoS, indicating that its determination
is important for our understanding of nuclear processes in dense matter.
 
The other important processes which intervene 
in accelerating a NS up to $P_{\rm{sh}}$ are the evolution of the mass transfer 
rate from the low mass companion star and the evolution of the NS magnetic moment 
$\mu$ \cite{5bpcdd99}. Within the recycling scenario \cite{5acrs82}\cite{5b95}
all schemes suggested for the origin and the evolution 
of $\mu$ allow the accreting NSs to reduce their $\mu$ and $P$ to the 
values that are characteristic of MSPs. However, due to the large volume of 
the parameters in the mass transfer scenario and to the strong 
coupling between it and the field evolution of the NS
in a low mass binary (LMB), only through a statistical approach it is possible to
establish how 
efficient the recycling process is in spinning a NS to 
$P<P_{\rm{min}}$. This was recognized first by Possenti {\it et al.} 
\cite{5pcdb98}\cite{5pcgbd99}, 
who carried on statistical analyses of NSs in the millisecond 
and sub-millisecond range, using a Monte Carlo 
population synthesis code with $\sim 3000$ particles.
It accounts for the evolution during the 
early phases when the NS in the LMB behaves as if isolated ($\e$ and $\p$) and later
when fed by wind accretion ($\ac$). The mass transfer during RLO
(again $\ac$) is modelled considering a range of accretion rates which 
is close to the one observed. The population synthesis 
code follows also the last radio-pulsar phase, whose duration is chosen from a 
flat probability distribution in the logarithm of time (see
Table~\ref{tab:input_values_pop} for a summary of the parameters
used in the population synthesis code). 
The model incorporates the detailed physics of the evolution of a 
crustal magnetic field (as discussed in $\S 3.3$, using BCI and BCII
to mimic expulsion 
or assimilation of the field in the NS core), and includes the relativistic 
corrections \cite{5bpcdd99} necessary to describe the spin-up process. 

\begin{table}
\caption{Population syntheses parameters}
\begin{center}
\renewcommand{\arraystretch}{1.4}
\setlength\tabcolsep{5pt}
\begin{tabular}{cccc}
\hline\noalign{\smallskip}
{Physical quantity}         & {Distribution} &              {Values }                             &   {Units}
\\
\hline
NS period at ${\rm t_{0}^{RLO}~(^*)}$  &     Flat     &                  1 $~~\to~~$ 100                      &   sec
\\
NS $\mu$ at ${\rm t_{0}^{RLO}~(^*)}$  &   Gaussian   & Log$<\mu_0>$=$$28.50$~;~\sigma$=0.32              &  G~cm$^{3}$
\\
${\dot{M}}$ in RLO phase$~(^{\sharp})$ &   Gaussian   &
Log$<{\dot{M}}>$=~1.00$~;~\sigma$=0.50       & $\rm{{\dot M}_{E}}$
\\
Minimum accreted mass                   &   One-value  &                       0.01                            & ${\rm {M_{\odot}}}$
\\
RLO accretion phase time~($^\dag$)     &  Flat in Log &
 10$^{6}~~\to~~\tau_{RLO}^{\rm{max}}$($^\ddag$) &    year
\\
MSP phase time                          &  Flat in Log &         
 10$^{8}~~\to~~$3$~\times~$10$^{9}$           &    year
\\
\hline
\hline
\end{tabular}
\end{center}
\footnotesize{
\noindent
{\bf (*)~} ${\rm t_{0}^{RLO}}$ = initial time of the Roche Lobe Overflow phase
\\
{\bf ($\sharp$)~} baryonic accretion rate during the Roche Lobe Overflow phase
\\
{\bf ($\dag$)~} a Maximum accreted Mass of $0.5 M_\odot$ is permitted 
during the RLO phase
\\
{\bf ($\ddag$)~} max duration of the RLO phase; 
explored values: 5$\times$10$^7$ yr - 10$^8$ yr - 5$\times$10$^8$ yr
}
\label{tab:input_values_pop}
\end{table}

Evolution is followed also beyond the RLO $\ac$ phase when accretion terminates.
The increasing evidence that NSs in LMBs may suffer phases 
of transient accretion (perhaps due to thermal-viscous instabilities 
in an irradiation dominated disc \cite{5v96}\cite{5kfkr97}) is suggestive 
that mass transfer onto a NS may not stop suddenly: the star probably 
undergoes a progressive reduction of the mean accretion rate, 
modulated by  phases of higher and lower accretion.
This in turn can start a cycle of $\p$ phases   
which in principle could vanish the effect of the previous  spin-up.
With the aim of exploring the effect of a decaying  $\dot M$ on the 
population of fastly spinning objects, two possibilities have been 
investigated: a persistent accretion for a time $\tau_{RLO}$ or
a persistent accretion for a shorter time followed
by a transient phase mimicking the quenching of the mass transfer.
The quenching of accretion has been modelled as a 
power law decay for $\dot M$ with index $\Gamma$ varying from $1 \to 10$ 
(the last value is representative of an almost sudden switch off).

\begin{figure}[htbp]
\begin{center}
\includegraphics[width=.45\textwidth]{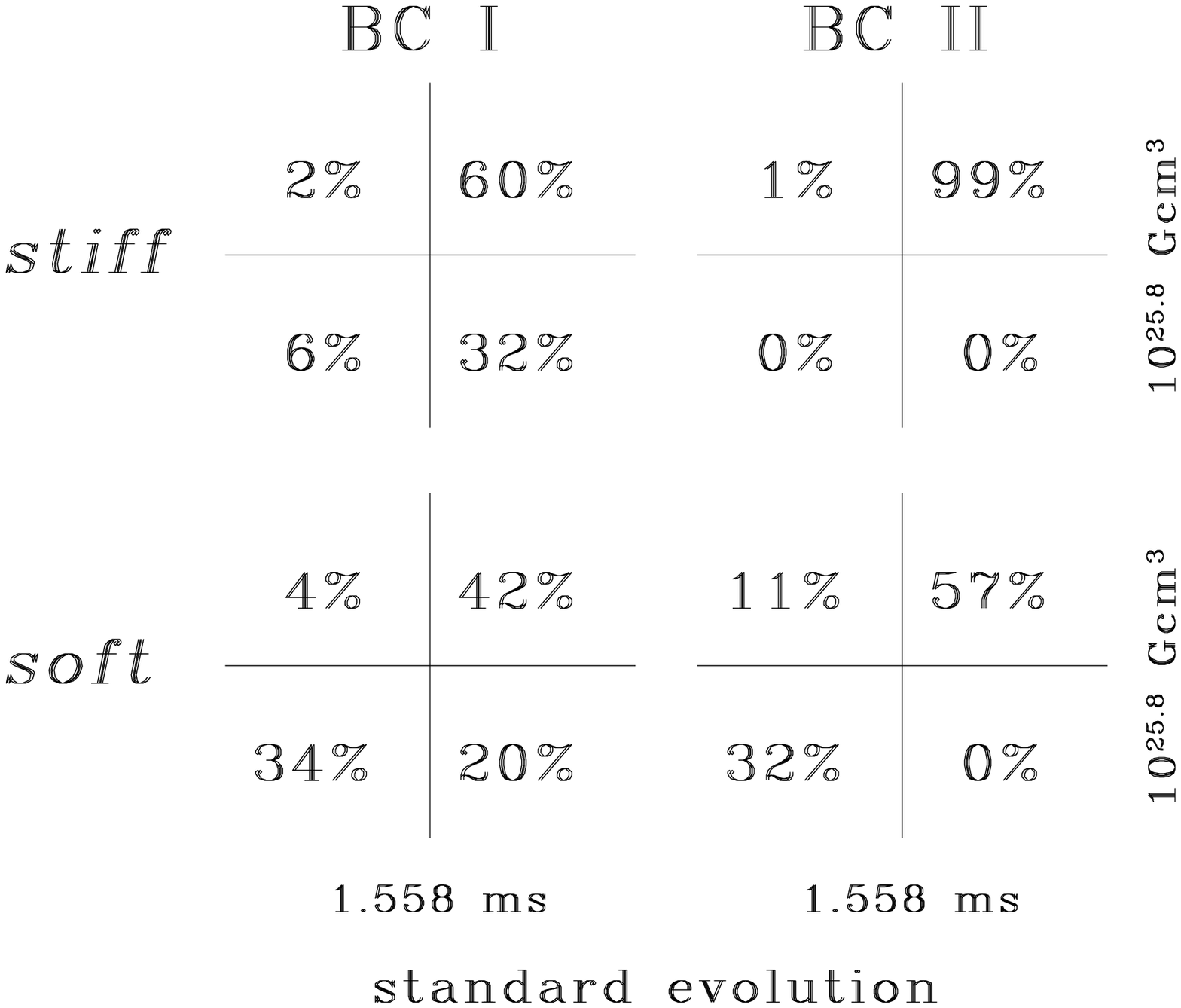}
\end{center}
\caption[]{Distributions of the synthesized NSs, 
derived normalizing the sample to the total number of model stars with 
$P<10~{\rm ms}$ and arbitrary value of  $\mu$.
The $\mu-P$ plane is divided in four regions.
As a guideline the upper left number in each cross gives the percentage of
objects having $P<P_{\rm{min}}$ and $\mu>\mu_{\rm{min}}$ (the typical variance is
about 1\%).}
\label{fig:Crosses}
\end{figure}

Figure~\ref{fig:Crosses} collects the fractional distribution of the
recycled model NSs using the set of parameters reported
in Table \ref{tab:input_values_pop} with $\tau_{RLO}^{{\rm max}}=5\times10^8$ yr.
Guided by  the values of $P_{\rm{min}}$ and $\mu_{{\rm min}}$ (the shortest 
rotational period observed in PSR~1937+21 and the weakest magnetic moment 
observed in PSRJ2317+1439), the particles are divided in four groups. 
Those filling the first quadrant ($P\geq P_{\rm{min}}$ and $\mu\geq\mu_{\rm{min}}$) 
behave as the known MSPs. Also the objects belonging to the second 
quadrant  ($P<P_{\rm{min}}$ and $\mu\geq\mu_{\rm{min}}$) should shine as PSRs
\cite{5bd97}. The effective observability of the objects 
in the third quadrant ($P<P_{\rm{min}}$ and $\mu < \mu_{\rm{min}}$) as radio sources 
represents instead a challenge for the modern pulsar surveys. Most of them 
will be above the ``death-line'' \cite{5cr93}, 
and might have a bolometric luminosity comparable to that of the known MSPs.  
Thereafter we shortly refer to {\it sub-}MSPs as to all  objects 
having $P<P_{\rm{min}}$ and $\mu$ above the ``death-line''.  
Objects in the fourth quadrant ($P\geq P_{\rm{min}}$ and $\mu < \mu_{\rm{min}}$)
are probably radio quiet NSs, because they tend to be closer
to the theoretical ``death-line'', and they are in a period range which
was already searched with good sensitivity by the radio surveys.

The four crosses (Fig.~\ref{fig:Crosses}) are for two representative EoSs (a very stiff
EoS with a
break-up period $1.4$ ms and a mildly soft one, with $P_{\rm{sh}}\simeq 0.7$ ms) 
and for two different boundary conditions (BCI \& BCII) for the magnetic 
field at the crust-core interface. 
It appears that objects 
with periods $P<P_{\rm{min}}$ are present in a statistically significant number. 
In effect, {\it a tail of potential sub-}MSPs {\it always emerges} for 
any reliable choice of the parameters listed in  
Table \ref{tab:input_values_pop}. 

\begin{figure}[htbp]
\begin{center}
\includegraphics[width=.5\textwidth]{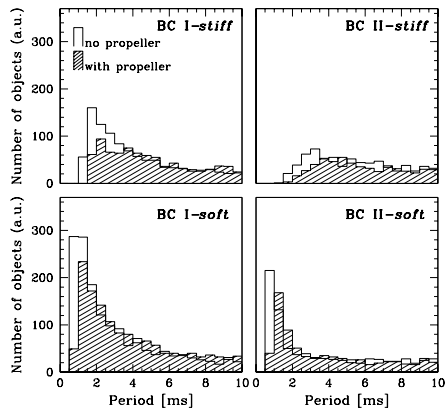}
\end{center}
\caption[]{Calculated distribution of millisecond NSs 
as a function of the spin period $P$ ($\mu$ is let vary over the whole range). 
{\it Solid line} denotes the distribution in absence of propeller, whilst 
{\it dashed area} denotes the case of a strong propeller effect. 
The absolute number of objects is in arbitrary units.}
\label{fig:Pdistribution}
\end{figure}

In the synthetic sample for the {\sl mild-soft} EoS, the ``barrier'' 
at $P_{\rm{sh}}\simeq 0.7$ ms is clearly visible in Figure ~\ref{fig:Pdistribution}
(solid lines): the {\it mild-soft EoS gives rise to period-distributions that 
increase rather steeply toward values smaller than 2} ms, irrespective to the 
adopted BCs. Instead, the boundary condition affects the distribution of $\mu$:
BCII produces a smaller number of objects with low field.
Thus, {\it sub}-MSPs could be a tool to test the physics of magnetic field decay.
\footnote{Note that a simple spin-down induced decay scenario has difficulty
in explaining the observed MSPs and predicts no ultra fastly spinning objects,
as shown in \cite{5pcdb98}.}

Even the {\sl very stiff} EoS permits periods $P<P_{\rm{min}}$, 
but the ``barrier'' of mass shedding (at $P_{\rm{sh}}\simeq 1.4$ ms) is so
close to $P_{\rm{min}}=1.558$ ms that only few NSs reach these extreme rotational 
rates. Moreover the period distribution for the stiff EoS is much flatter 
than that for the soft-EoS, displaying a broad maximum at $P\sim 3$ ms.
It has been recently claimed that X-ray sources in LMBs show rotational 
periods clustering in the interval $2 \to 4$ ms \cite{5wz97} \cite{5v98}. 
This effect could be explained introducing a fine tuned relation 
between $\mu$ and $\dot M$ ($\mu \propto {\dot{M}}^{1/2}$ \cite{5wz97}). 
Alternatively, 
gravitational wave emission has been invoked \cite{5aks99}\cite{5b98}. 
Here we notice that such a clustering can be a natural statistical 
outcome of the recycling process if the EoS for the nuclear 
matter is stiff enough.

Physical ingredients to describe the propeller induced 
spin-down of a NS at the end of the RLO phase are poorly known or difficult 
to assess (e.g. the exact law for
the decrease of the mass transfer rate or the efficiency in the
extraction of the angular momentum from the NS to the propelled matter). 
In our scheme, the largest 
effect occurs when phase $\p$ lasts half  ($\sim50\%$) of the
RLO phase and the power-law index is $\Gamma\sim8$. Figure ~\ref{fig:Pdistribution} 
reports the period distributions (dashed areas) when such a strong 
propeller phase is included. We note that {\it a strong propeller can
threaten the formation of NSs with $P<P_{\rm{min}}$ and $\mu>\mu_{\rm{min}}$}
in the case  of the very stiff EoS, whilst {\it for the mild-soft EoS the 
distributions preserve a maximum just about $P_{\rm{min}}.$}

\begin{figure}[htbp]
\begin{center}
\includegraphics[width=.55\textwidth]{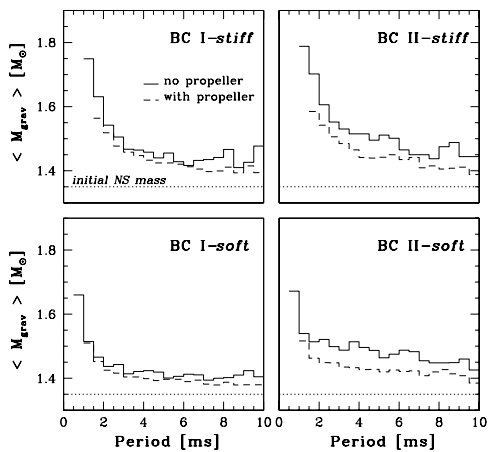}
\end{center}
\caption[]{Average gravitational masses of the re-accelerated NSs 
as a function of their final spin period $P.$ $\mu$ varies over the whole 
range and the synthesized NSs are binned in $0.5$ ms wide intervals.
The initial mass of the static NSs is set equal to $1.35 {\rm M_{\odot}}$ in\
all the cases. {\it Thick solid line} denotes the mass-distribution in 
absence of propeller, whereas {\it thin dashed line}  with a 
strong propeller included.}
\label{fig:Mdistribution}
\end{figure}

This statistical analysis provides also information on the NS mass 
distribution as a function of $P$ at the end of evolution. 
The initial NS gravitational mass, in the evolutionary code, was set at 
$M_{G}=1.35~{\rm M_\odot}$ (according to the narrow Gaussian
distribution with $\sigma=0.04~{\rm M_\odot}$ resulting from 
measuring the mass of the NSs in five relativistic NS-NS binary systems
\cite{5tc99}). Figure \ref{fig:Mdistribution} clearly shows that 
the observed millisecond population should have undergone a 
mass load of $\lsim 0.1 {\rm M_\odot}$ during recycling.
This is consistent with the few estimates of the masses of millisecond pulsars 
in low mass binaries \cite{5nss01} and raises the problem of explaining
how the NS can get rid of the bulk of the mass ($0.5-1.5~{\rm M_\odot}$) 
released by the companion during the RLO  phase \cite{4ts99}.

\begin{figure}[htbp]
\begin{center}
\includegraphics[width=.49\textwidth]{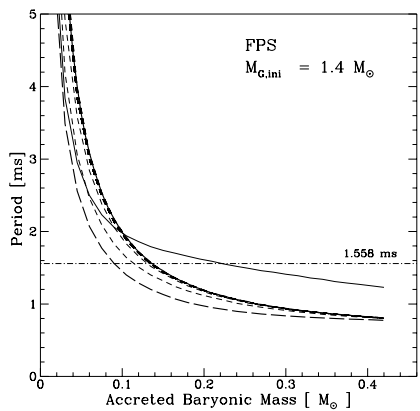}
\end{center}
\caption[] 
{Period versus Accreted Baryonic Mass for a magnetized NS using FPS-EoS and 
an initial gravitational mass of $1.40~{\rm {M_\odot}}$. The different pathways 
(a sample is represented by the {\it dashed lines}) define a strip, which 
narrows towards shorter periods. The strip is upper 
bounded by the evolutionary path for an unmagnetized NS ({\it bold solid 
line}). The {\it bold long dashed line} refers to the evolution along the
spin-up line and is calculated assuming a tuned torque function 
(which maximizes the efficiency of the 
spin-up process). The {\it thin solid line} refers to a very slow
decay of $\mu$, implying larger mass depositions in order to attain very short 
spin periods.
}
\label{MinMacc}
\end{figure}

Figure \ref{fig:Mdistribution} suggests that the mass function steepens 
toward high values only when $P$ falls below $\sim P_{\rm{min}}$,
approaching $M\sim 1.7 \div 1.8~{\rm{M_{\odot}}}$. That is a straight consequence 
of a results already pointed out by Burderi {\it et al.} \cite{5bpcdd99}: 
a large mass deposition (at least $\gsim 0.25 {\rm M_\odot}$) is required 
to spin a NS to ultra short periods, as illustrated in Figure \ref{MinMacc}. 
The action of the propeller during 
the evolution has 
the effect of reducing the mass infall: the mass distribution 
is only slightly affected for the mild-soft EoS, while for the stiff EoS 
the difference is more pronounced.

The wandering among the various phases can account for the interesting
phenomenology of the MSPs observed and opens the possibility that even more
extreme objects like the {\it sub-}MSPs exist in the Galaxy.

\section{The NS census in the Milky Way}
The isolated NSs follow a completely different evolution pattern. 
Generally, 
over the Hubble time, they cover large portion of the Milky
Way and thus explore  regions where the ISM is inhomogeneous 
looping  among the various phases erratically.
With a population synthesis model,  
Popov {\it et al.}\cite{p2000}  traced their evolution in the ISM, and in the Milky Way
potential.
The model NSs, born in the galactic plane at a rate proportional to the square of
the
ISM
density, have initially
short spin periods, magnetic fields clustered around $10^{30}$ G cm$^{3}$, and 
spatial (kick) velocities that can be drawn for a Maxwellian distribution 
(with mean velocity modulus $\langle V \rangle$ treated as a parameter).

\begin{figure}[htbp]
\begin{center}
\includegraphics[width=.6\textwidth,angle=-90]{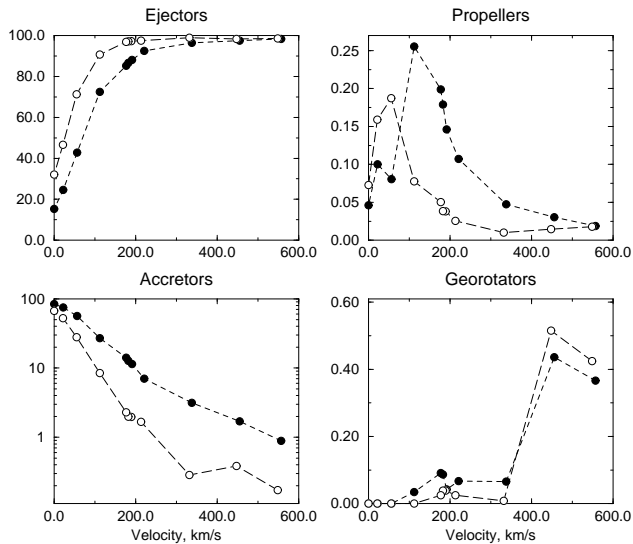}
\end{center}
\caption{Fractions of NSs (in percents)    
in phases $\e,\p\ac$ and $\g$  versus the mean kick      
velocity $\langle V\rangle,$ for a constant 
magnetic moment $\mu=0.5\times 10^{30} {\rm G}\, {\rm
cm}^3$ (open circles) 
and $\mu=10^{30} {\rm G}\, {\rm cm}^3$
(filled circles); typical statistical uncertainty for $\e$ and $\ac$ 
$\sim $ 1-2\%.}
\label{census_iso}
\end{figure}

The collective properties are illustrated in Figure \ref{census_iso} for two initial
values of $\mu_0=0.5-1\times 10^{30}$ G cm$^3$:
For a non evolving field most of the isolated NSs spend their lives
as Ejectors and there is no possibility to observe them after the radio pulsar phase.
Propellers are shortlived \cite{lp95}, and  Georotators are  rare.
A tiny fraction (a few percents) on the NSs are in the Accretor stage if 
$\langle V\rangle$ is above 200 km s$^{-1}.$  Only in the unrealistic case of a low
mean velocity, the bulk of the population would be in stage $\ac$.

\begin{figure}[htbp]
\begin{center}
\includegraphics[width=.5\textwidth,angle=-90]{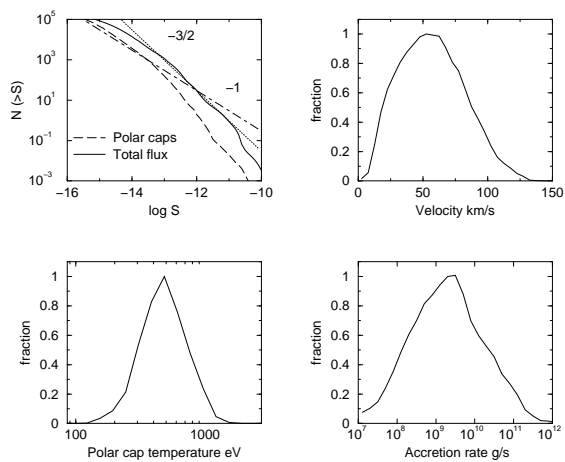}
\end{center}
\caption{
Upper left panel:  the log N-log S distribution for Accretors within 5 kpc from the
Sun. Dashed (solid) curve refers to emission from polar cap (entire NS surface)
in the range 0.5-2.4 keV:
straight lines with slopes -1 and -3/2
 are included for comparison. Fro top right to bottom right: the velocity $V$,
effective polar cap temperature and accretion rate $\dot M$ distributions for
Accretors \cite{6pop00}}
\label{accretion_iso}
\end{figure}

As illustrated in Figure \ref{accretion_iso},
in phase $\ac,$ the mean
value of $\dot {M}$ is a few $10^9$
g s$^{-1}$ and,
among the Accretors, the typical velocity clusters around $50$ km s$^{-1},$ 
and the  luminosity $L\sim (GM/R)\dot {M}$
around  $10^{29}$ erg s$^{-1}.$  Accreting isolated NSs are  "visible", 
but they would be extremely dim
objects
emitting predominantly in the soft X-rays, with a polar cap black body 
effective temperature of 0.6 keV.
What can we learn when comparing the
theoretical census with the observations?
We can discover if  field decays in isolated objects, if the population is devoid of 
slow objects, and if the stars effectively spin  
down due to the unavoidable interaction with the tenuous ISM.
The search of accreting isolated NSs is thus compelling
\cite{6ors70}\cite{6tc91}\cite{6bm93}\cite{4ttzc00}.

\subsection{Accreting isolated neutron stars in the {\sl Rosat} Sky?}

Despite intensive observational campaigns, no irrefutable identification
of an isolated accreting NS has been presented so far.   
Six soft sources have been found in {\sl Rosat} field \cite{4ttzc00},
identified as isolated NSs from the optical and X-ray data.
Observations, however, do not permit to unveil the origin of their emission, yet.
These sources could be  powered either by accretion 
or by the release of internal energy
in relatively young ($\approx 10^6$ yr) 
cooling NSs \footnote {see \cite{4ttzc00} for an updated review, the description of
the
sources and a complete reference list.}.
Although relatively bright 
(up to $\approx 1 \ {\rm cts\,s}^{-1}$), 
the proximity of the sources (inferred from the column density in the X-ray spectra)
makes their luminosity  
($L\approx 10^{31} \ {\rm erg\, s}^{-1}$)
near to that expected from either a close-by cooling NS or
from an accreting NS, among the most luminous.
Their X-ray spectrum
is soft and thermal, again as predicted for both Accretors and Coolers.

Can a comparison with theoretical expectations help in discriminating among the two
hypothesis?
First, the paucity of very soft X-ray sources
in the {\sl Rosat} fields  (in comparison with earlier expectation
\cite{6tc91}\cite{6bm93}) is indicating that Accretors are rare objects.
If these six sources are indeed accreting, 
this implies that the average velocity of the isolated 
NSs population at birth has to
exceed $\sim 200 \ {\rm km\, s^{-1}}$ (a figure
consistent with that derived from radio pulsars statistics \cite{6ll94}). 
In addition, since observable accretion--powered isolated NSs are (intrinsically) slow
objects, 
these results exclude the possibility that the present velocity
distribution of NSs is richer in low--velocity objects with   
respect to a Maxwellian. 
Thus, despite the fact that in our Galaxy there are many old NSs (about $10^9$), 
the young ($\sim 10^6$ yr) cooling NSs seem to outnumber, at the {\sl Rosat} counts,
those
in phase $\ac$.
This is what emerges also from the calculation of the log N-log S distribution
both of cooling and accreting stars (the last carried on with "census";
\cite{6pop00}).
At the bright counts, the local population of cooling NSs would
dominate over the
log N-log S of the dimmer and warmer (less absorbed) Accretors. There is the hope that  
{\sl Chandra}
and {\sl Newton}  will detect them, at the flux
limit of $10^{-16}$  erg cm${^{-2}}$ s$^{-1}$ \cite{6pop00}.
In support of the cooling hypothesis there is also  
the recent measurement\cite{6wm97}\cite{6w00}  of the velocity
($200$ km s$^{-1}$) of RXJ1856 (a member of this class) indicating that this
source is, most likely, not powered by accretion having a high velocity. 
Interestingly, the cooling hypothesis 
(when explored in the log N-log S plane) implies that  the NS birth rate in the
Solar vicinity, over the last 10$^6$ yr, is higher than that inferred from radiopulsar
observations \cite{6pop00} \cite{6nt99}. This might be a crucial point as it may
indicate that most of the 
young NSs may not share the observational properties of the canonical PSRs,
hinting for the presence of a background of "anomalous" young cooling NSs. 

What about field decay in the accretion hypothesis? Decay of the field to
extreme low values would imply a large number of
Accretors \cite{p2000}, not observed, thus excluding this possibility. 
This indicates that, in crustal models \cite{3sc87} \cite{3um92}, electric currents 
need to be located deep into the
crust and that the impurity parameter needs to be not exceedingly large \cite{3kb99},
in spin-down induced models.
What is still uncertain is whether the paucity is due mainly
to a velocity effect than to a "fine-tuned" decay, a problem not solved yet,
statistically.  

While our theoretical expectation hints in favor of the cooling hypothesis
there remain yet a puzzle:  the long period of one of these sources,
RXJ0720 \cite{6haberl97}.
If powered by accretion, RXJ0720 would have a "weak" field NS ($\mu\gsim 10^{26}$ G
cm$^{3}$) and
this would be a rather 
direct prove of some field decay, in isolated NSs\cite{6kp97}\cite{6w97}. 
Can a young cooling object have such a slow rotation?  
Do we have to change or view that NSs come to birth with ultra short periods?
A new challenging hypothesis has been put forward \cite{6haberl97}\cite{6hk98}
\cite{6cgp00}
that
RXJ0720 is
just the descendant
of a highly magnetized NS (a {\it Magnetar}) that have suffered a severe spin down
accompanied by a nonlinear decay of the field whose energy is powering the X-ray
luminosity \cite{6td96}.
This issue remains one of the new problems of the NS physics, making
the debate on the nature of these sources an even more exciting problem.

\section{Conclusions}

In this review we have traced the evolution of old NSs transiting through 
the Ejector, Propeller and Accretor phases.
The NSs of our review are far from being 
"canonical" MSPs
in light binaries, or 
"canonical" PSRs in the field.
The ultra fastly spinning NSs, that we have recycled in binaries,
are rather extreme relativistic  heavy NSs:
If discovered, they will unable us to probe the stellar
interior in an unprecedented way, and to constrain the physics driving magnetic
field decay, in interacting systems.
As regard to the field NSs, the discovery of the six {\sl Rosat} sources
has just opened the possibility of unveiling "unconventional"  NSs,
evolving in isolation. Whether they are accreting or cooling objects is still a
mystery and even more mysterious and fascinating is their possible link with 
"Magnetars".
The study of these "unusual" NSs can open new frontiers in this already active
field of research.

\section{Acknowlegdments}

The authors would like to thank the European Center for Theoretical Physics
and the organizers 
of the conference ``Physics of
Neutron Star Interiors'', D. Blaschke, N.K. Glendenning and A. Sedrakian, 
for kind hospitality during the workshop.

\end{document}